\documentclass[
preprint,
sort&compress,
times,
lefttitle
]{elsarticle}

\usepackage{amssymb}
\usepackage{amsmath}
\usepackage{siunitx}
\usepackage{pifont}
\usepackage{multirow}
\usepackage{color}

\usepackage{array}
\usepackage{lineno}
\usepackage{threeparttable}
\usepackage{nicefrac}

\usepackage{hyperref}
\hypersetup{colorlinks=true, breaklinks=True}
\usepackage{graphicx}

\makeatletter

\newcommand{\Rmnum}[1]{\expandafter\@slowromancap\romannumeral #1@}
\makeatother

\newcommand{\nuc}[2]{${}^{#1}\mathrm{#2}$}
\newcommand{\ee}[1]{\times 10^{#1}}
\newcommand{\etal}[1]{#1 \textit{et al.}}
\newcommand{\Nean}{${}^{22}\mathrm{Ne}(\alpha,n){}^{25}\mathrm{Mg}$}
\newcommand{\Neag}{${}^{22}\mathrm{Ne}(\alpha,\gamma){}^{26}\mathrm{Mg}$}

\newcommand{\revised}[1]{#1}

\begin{document}
%

\title{Decay properties of $^{22}\mathrm{Ne} + \alpha$ resonances and their impact on  $s$-process nucleosynthesis}

\author[1]{S. Ota}\corref{cor1}
\ead{shuyaota@comp.tamu.edu}
\author[1,2,3]{G. Christian\corref{cor1}\fnref{fn}}
\ead{gchristian@tamu.edu}
\author[4]{G. Lotay}
\author[4]{W. N. Catford}
\author[1,2]{E. A. Bennett}
\author[1,2]{S. Dede}
\author[4]{D. T. Doherty}
\author[4]{S. Hallam}
\author[1,2]{J. Hooker}
\author[1,2]{C. Hunt}
\author[1,2]{H. Jayatissa}
\author[4]{A. Matta\fnref{fn2}}
\author[4]{M. Moukaddam\fnref{fn3}}
\author[1,2,3]{G. V. Rogachev}
\author[1]{A. Saastamoinen}
\author[4]{J. A. Tostevin}
\author[1,2]{S. Upadhyayula}
\author[4]{R. Wilkinson}

\cortext[cor1]{Corresponding author}
\fntext[fn1]{Present address: Department of Astronomy \& Physics, Saint Mary's University, Halifax, NS B3H 3C3 Canada}
\fntext[fn2]{Present address: Laboratoire de Physique Corpusculaire de Caen, 14050 CAEN CEDEX 4, France}
\fntext[fn3]{Present address: PHC-DRS/Universit{\'e} de Strasbourg, IN2P3-CNRS, UMR 7178, F-67037, Strasbourg, France.}

\address[1]{Cyclotron Institute, Texas A\&M University, College Station, TX 77843, USA}
\address[2]{Department of Physics \& Astronomy, Texas A\&M University, College Station, TX 77843, USA}
\address[3]{Nuclear Solutions Institute, Texas A\&M University, College Station, TX 77843, USA}
\address[4]{Department of Physics, University of Surrey, Guildford GU2 7XH, UK}

\date{\today}
 
\begin{abstract}
The astrophysical $s$-process is one of the two main processes forming  elements heavier than iron. A key outstanding uncertainty surrounding $s$-process nucleosynthesis is the neutron flux generated by the ${}^{22}\mathrm{Ne}(\alpha,  n){}^{25}\mathrm{Mg}$ reaction during the He-core and C-shell burning phases of massive stars. This reaction, as well as the competing ${}^{22}\mathrm{Ne}(\alpha, \gamma){}^{26}\mathrm{Mg}$ reaction, is not well constrained in the important temperature regime from ${\sim} 0.2$--$0.4$~GK, owing to uncertainties in the nuclear properties of resonances lying within the Gamow window. To address these uncertainties, we have performed a new measurement of the ${}^{22}\mathrm{Ne}({}^{6}\mathrm{Li}, d){}^{26}\mathrm{Mg}$ reaction in inverse kinematics, detecting the outgoing deuterons and ${}^{25,26}\mathrm{Mg}$ recoils in coincidence. We have established a new $n / \gamma$ decay branching ratio of $1.14(26)$ for the key $E_x = 11.32$~MeV resonance  in $^{26}\mathrm{Mg}$, which results in a new $(\alpha, n)$ strength for this resonance of $42(11)~\mu$eV when combined with the well-established $(\alpha, \gamma)$ strength of this resonance.  We have also determined new upper limits on the $\alpha$ partial widths of neutron-unbound resonances at $E_x = 11.112,$ $11.163$, $11.169$, and $11.171$~MeV.  Monte-Carlo calculations of the stellar ${}^{22}\mathrm{Ne}(\alpha, n){}^{25}\mathrm{Mg}$ and ${}^{22}\mathrm{Ne}(\alpha, \gamma){}^{26}\mathrm{Mg}$ rates, which incorporate these results, indicate that both  rates are substantially lower than previously thought in the temperature range from ${\sim} 0.2$--$0.4$~GK.
\end{abstract}


\maketitle


\section{Introduction}

Understanding the production of nuclides heavier than iron is a crucial part of our global quest to understand the origin of the elements. The slow neutron capture process ($s$-process) is a key contributor to heavy-element synthesis, producing around $50\%$ of the nuclides heavier than iron in our solar system. The $s$-process occurs in relatively moderate stellar environments---the He-shell burning phase of intermediate-mass asymptotic giant branch (AGB) stars and the He-core and C-shell burning phases of massive ($M > 8M_{\odot}$) stars---and involves a series of neutron capture reactions on stable or near-stable nuclei \cite{Kappeler2011}. Neutron capture rates on stable isotopes are typically known to an accuracy of $20\%$ or better, and hence the $s$-process offers a prime opportunity to compare predicted nucleosynthesis yields with astronomical observations and meteorite and stardust analyses \cite{Kappeler2011,Karakas2012,Clayton2004,Lugaro2005}. However, there are still a number of key outstanding nuclear physics uncertainties surrounding the $s$-process---in particular, neutron capture rates on branching-point nuclides, as well as uncertainties related to the overall neutron flux. Neutron generation during the weak $s$-process, which occurs during the He-core and C-shell burning phases in massive stars, is dominated by the $^{22}\mathrm{Ne}(\alpha,n){}^{25}\mathrm{Mg}$ reaction. The rate of this reaction, as well as the competing $^{22}\mathrm{Ne}(\alpha,\gamma){}^{26}\mathrm{Mg}$ reaction, is not well constrained in the relevant temperature range for the weak $s$-process, significantly impacting predicted nucleosynthesis yields.
For example, recent calculations indicate that varying the $^{22}$Ne($\alpha,n$)$^{25}$Mg rate within existing uncertainties leads to a factor of ten or greater changes in predicted weak $s$-process yields throughout the $A \sim 60$--$90$ mass region \cite{Longland2012,Talwar2016}.

The astrophysical impact of the $^{22}$Ne($\alpha,\linebreak[1] n$)$^{25}$Mg and $^{22}$Ne($\alpha,\linebreak[1]  \gamma$)$^{25}$Mg reactions is not limited to the $s$-process. They also affect synthesis of the long-lived $\gamma$-ray emitters such as $^{60}$Fe, created through ${}^{59}$Fe$(n,\gamma){}^{60}$Fe with $^{22}$Ne($\alpha,n$)$^{25}$Mg serving as a neutron source in the high-temperature ($T \sim 1$~GK) C-shell burning phase of massive stars. The $\gamma$-ray emission from $^{60}$Fe has been observed together with $^{26}$Al, another long-lived $\gamma$-ray emitter, by low-energy $\gamma$-ray telescopes such as INTEGRAL \cite{Wang2007}. Their abundance ratio is considered a key constraint on massive-star nucleosynthesis models and galactic chemical evolution \cite{Diel2010}. Furthermore, estimated chemical abundances based on established $s$-process models provide estimates of the less well-known solar $r$-process abundances. Hence a better understanding of $s$-process nucleosynthesis is helpful in working towards a complete understanding of the $r$-process in the era of multi-messenger astronomy.

At stellar temperatures, both $^{22}\mathrm{Ne}(\alpha,\linebreak[1] n){}^{25}\mathrm{Mg}$ and $^{22}\mathrm{Ne}(\alpha,\linebreak[1] \gamma){}^{26}\mathrm{Mg}$ proceed through resonant capture to natural-parity states in the compound nucleus, $^{26}\mathrm{Mg}$. For a given resonance, the key properties determining its contribution to the stellar rate are the $\alpha + {}^{22}\mathrm{Ne}$ resonance energy and the resonance strengths, $\omega\gamma_{(\alpha,n)}$ and $\omega\gamma_{(\alpha,\gamma)}$.\footnote{The resonance strengths are given by $\omega\gamma_{(\alpha,n)} \simeq (2J+1)\Gamma_{\alpha}/(1+\Gamma_{\gamma}/\Gamma_n$) and $\omega\gamma_{(\alpha,\gamma)} \simeq (2J+1)\Gamma_{\alpha}/(1+\Gamma_{n}/\Gamma_\gamma$), taking the approximation $\Gamma_\alpha \ll \Gamma$. Here, $J$ is the resonance spin, and $\Gamma_\alpha$ ($\Gamma_n,$ $\Gamma_{\gamma}$, $\Gamma$) is the $\alpha$ (neutron, $\gamma$-ray, total) partial width.} 
Above $0.3$~GK, both the $(\alpha,n)$ and $(\alpha,\gamma)$ reactions are dominated by a resonance at $E_{cm} = 0.703$~MeV ($E_x=11.32$~MeV). This resonance has been observed in both direct $(\alpha,n)$ \cite{Harms1991, Drotleff1993, Giesen1993, Jaeger2001} and $(\alpha,\gamma)$ experiments \cite{Wolke1989,Hunt2019}, although there is disagreement about whether or not these are the same state \cite{Koehler2002, Longland2012, Adsley2017}.
The $(\alpha,\gamma)$ strength is well established, with both published measurements \cite{Wolke1989, Hunt2019} in good agreement, as well as with the unpublished result of Jaeger \cite{JaegerPhD}. In contrast, the resonance strengths extracted from the direct $(\alpha,n)$ measurements are in poor agreement, suggesting the presence of an unknown systematic bias in the data \cite{Longland2012}. As a result, $\omega\gamma_{(\alpha,n)}$ for the $E_x=11.32$~MeV  resonance remains a key outstanding uncertainty on the total $^{22}\mathrm{Ne}(\alpha,n){}^{25}\mathrm{Mg}$ rate at stellar temperatures. Additionally, the spin-parity of this resonance is not firmly established, and most recently it was suggested that $J^\pi = (0^+,$ $1^-$, $2^+$, $3^-$) are all allowable \cite{Hunt2019}. Although $J^\pi$ does not affect the direct-measurement strengths, it is a crucial parameter for extracting $\omega\gamma_{(\alpha,n)}$ from indirect $\alpha$ transfer studies \cite{Giesen1993, Talwar2016}. Additionally, it affects the scaling factors used to extract $\Gamma_\alpha$ for lower-lying resonances from $\alpha$ transfer \cite{Talwar2016}.

\revised{
At lower temperatures, both reactions \emph{may} be dominated by one or more resonances between the neutron threshold and $E_{cm} \sim 0.635$~MeV ($E_x \sim 11.25$~MeV). The presence of an important resonance in this region has long been controversial. \etal{Giesen}, in their  (\nuc{6}{Li},~$d$) experiment, observed no strong transitions in this energy region \cite{Giesen1993}. They set an upper limit of $S_\alpha < 0.02$ on a candidate $1^-$ resonance at $E_x=11.15$~MeV originally identified in  photoneutron \cite{Berman1969} and neutron capture \cite{Weigmann1976} studies. Assuming $J=1$ and contemporary values for $\Gamma_n$ and $\Gamma_\gamma$, this resulted in respective limits on the $(\alpha,n)$ and $(\alpha,\gamma)$ strengths of $\omega\gamma_{(\alpha,\gamma)} < 0.097~\mu$eV and $\omega\gamma_{(\alpha,n)} < 0.74~\mu$eV.
 \etal{Jaeger} also searched for this state in their direct $(\alpha,n)$ measurement, setting an upper limit on the resonance strength of $\omega\gamma_{\alpha,n} < 60~$neV. This state was later shown to be a $1^+$ (non-natural parity) and thus completely inconsequential to either $\alpha$ capture reaction \cite{Longland2009}.

In the later (\nuc{6}{Li},~$d$) measurement at higher beam energy, Talwar \textit{et al}.\ observed a strong transition at $E_x = 11.167(8)$~MeV \cite{Talwar2016}. They also observed a transition at a similar energy in their concurrent $^{26}\mathrm{Mg}(\alpha,\alpha^\prime){}^{26}\mathrm{Mg}$ experiment. They assigned a spin-parity of  $J^\pi = (1^-, 2^+)$ and extracted a spectroscopic factor of  $S_\alpha=0.36$, corresponding to $\Gamma_\alpha = 0.18(2)~\mu$eV when taking their preferred spin-parity assignments. In order to maintain consistency with the $\omega\gamma_{(\alpha,n)}$ upper limit of \etal{Jaeger}, they attributed the additional strength to the $(\alpha,\gamma)$ channel, establishing $\omega\gamma_{(\alpha,\gamma)} = 0.54(7)~\mu$eV. Based on these results, the authors established that this state dominates the \Neag{} rate between ${\sim}0.2$--$0.4~$GK and that it could potentially dominate the \Nean{} rate below $0.2$~GK.
More recently, neutron capture studies by Massimi \emph{et al}.\ have identified four natural-parity resonances in the ${}^{26}\mathrm{Mg}$ excitation energy range from $11.1$--$11.3$~MeV \cite{Massimi2012, Massimi2017}. Three of these resonances were found to stongly neutron decay, while the fourth, $E_x = 11.171$~MeV, was identified as a $2^+$ state with a significant $\gamma$-ray decay branch ($\Gamma_{n}/\Gamma_\gamma = 0.2$--$6$). This makes it a candidate for the strong $(\alpha,\gamma)$ resonance claimed in Ref.~\cite{Talwar2016}.
} 

Despite extensive investigaton \cite{Wolke1989,Harms1991, Drotleff1993, Giesen1993, Jaeger2001,Koehler2002, Ugalde2007, Schwengner2009, Longland2009, deBoer2010, Massimi2012, Ota2012, deBoer2014, Ota2014, Talwar2016, Massimi2017, Adsley2017, Adsley2018, Hunt2019, Longland2012, JaegerPhD, Berman1969, Weigmann1976}, key properties of both the $E_x = \linebreak[1] 11.32$ MeV resonance and lower-lying neutron-unbound resonances remain uncertain. In particular, the $(\alpha,n)$ strength of the  $E_x = 11.32$~MeV resonance is not well established, nor is its spin-parity. 
The situation for lower-lying resonances is even less clear, with substantial disagreement between existing $({}^6\mathrm{Li},d)$ experiments about the presence of a strong $\alpha$-cluster state around $E_x = 11.17$~MeV. Additionally, the $\alpha$ partial widths for the natural-parity resonances identified by Massimi \emph{et al}.\ remain poorly constrained, with the lowest-lying resonance, $E_x = 11.112$~MeV, having the potential to completely dominate the $(\alpha,n)$ rate below ${\sim}0.3$~GK.

Here we report a new, kinematically-complete measurement of the ${}^{22}\mathrm{Ne}({}^6\mathrm{Li}, d){}^{26}\mathrm{Mg}$ reaction in inverse kinematics, with direct sensitivity to the decay modes of observed states. The sensitivity to decay modes is a key advantage of the present study, allowing us to simultaneously address a number of questions surrounding both the $^{22}\mathrm{Ne}(\alpha,n){}^{25}\mathrm{Mg}$ and $^{22}\mathrm{Ne}(\alpha,\gamma){}^{26}\mathrm{Mg}$ reactions. In particular, we report a new value of $\Gamma_n / \Gamma_{\gamma}$ for the $E_x = 11.32$~MeV resonance that is a factor ${\sim}3$ below the value extracted from direct measurements.
Furthermore, in the region between the neutron threshold and $E_x = 11.25$~MeV, the sensitivity to decay modes allows us to set stringent limits on the ${}^{22}\mathrm{Ne}({}^6\mathrm{Li}, dn){}^{25}\mathrm{Mg}$ cross section, which translates into an upper limit on $\Gamma_{\alpha} \Gamma_{n}/\Gamma$. Taken in conjunction with the resonance parameters ($J$, $\Gamma_{n}$, $\Gamma_{\gamma}$) established by Massimi \emph{et al}., this results in new upper limits on both the $(\alpha,n)$ and $(\alpha,\gamma)$ strengths for the four natural-parity states in this region identified in Refs.~\cite{Massimi2012, Massimi2017}.

\section{Methodology and Results}
\begin{figure}
	\centering
 \includegraphics[width=8cm]{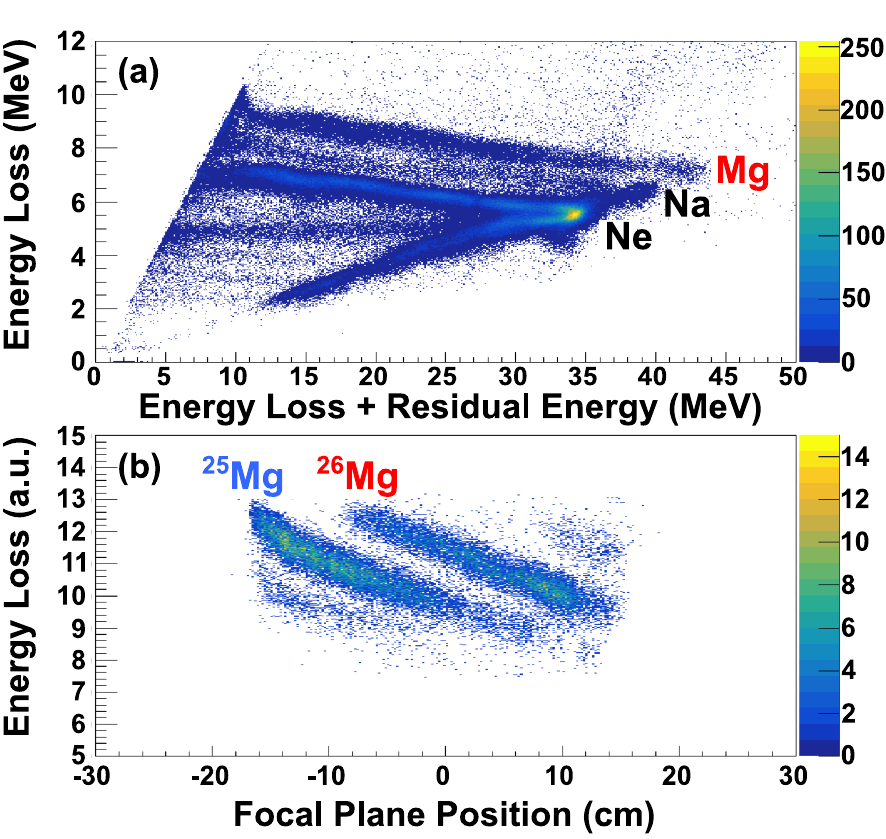}
 \caption{a) Energy loss vs.\ total energy measured in the Oxford detector, showing separation of Mg isotopes from other species. b) Energy loss vs.\ focal plane position, gated on the Mg band in panel (a) and showing clear separation of \nuc{25}{Mg} and \nuc{26}{Mg}.\label{fig:PID}}
 \end{figure}

The experiment was performed at the Texas A\&M University Cycloton Institute, using the K150 cyclotron to deliver a beam of 154~MeV $^{22}$Ne$^{(7+)}$ ions, impinging on a 30~$\mu$g/cm$^2$ $^{6}$LiF target (95\% \nuc{6}{Li} purity), with a 10 $\mu$g/cm$^2$ carbon backing. Our detector system consisted of the TIARA Si array \cite{Labiche2010}, four closely-packed HPGe clovers \cite{Lesher2010}, and the MDM spectrometer \cite{Pringle1986}. The acceptance of the MDM was $\pm 2^\circ$ in both the dispersive and non-dispersive planes, defined by rectangular slits at its entrance.
Target-like deuterons were detected in the backward hemisphere ($\theta_{lab}$ = 148$^\circ$--168$^\circ$) by a double-sided annular Si detector and were used to reconstruct the excitation energy of $^{26}$Mg states from the missing mass. Elastically scattered target nuclei were detected in a series of resistive strip detectors in a barrel configuration ($\theta_{lab}$ = 45$^\circ$--145$^\circ$). 
Beam-like \nuc{26}{Mg} (\nuc{25}{Mg}) recoils resulting from $\gamma$-ray (neutron) decay of $^{26}$Mg excited states were unambiguously identified in the MDM focal plane using a combination of energy loss, total energy, and dispersive position signals from the upgraded Oxford detector \cite{Spiridon2016,Spiridon2019}, as demonstrated in Figure~\ref{fig:PID}.  The identification of ${}^{25,26}$Mg recoils was confirmed by the coincident $\gamma$-ray transitions measured in the HPGe detectors. The clear identification of both recoil species with good efficiency allowed reliable extraction of $\Gamma_n/\Gamma_\gamma$ for $^{26}$Mg states populated in the $({}^6\mathrm{Li},d)$ reaction, from the efficiency-corrected ratio of ${}^{25}\mathrm{Mg}/{}^{26}\mathrm{Mg}$ recoils in coincidence with the state of interest.


\revised{
These recoil detection efficiencies are the product of the spectrometer acceptance, the Mg charge state fraction, and the intrinsic detection efficiency of the focal plane detectors. The charge state fraction is identical for \nuc{25,26}{Mg} and hence cancels in the final ratio used to calculate $\Gamma_n/\Gamma_\gamma$. The intrinsic detection efficiency was estimated to be $80.0(20)\%$. This was calculated from the ratio of ${}^{23}\mathrm{Ne} + d$ coincidences to deuteron singles observed for strongly-populated singlet states in a separate ${}^{22}\mathrm{Ne}(d,p){}^{23}\mathrm{Ne}$ run using the same setup. Due to the small cone angle of the $(d,p)$ reaction,  the \nuc{23}{Ne} acceptance here was $100\%$. The intrinsic efficiency again cancels in the $\Gamma_n/\Gamma_\gamma$ calculation; however, the $\pm 2.0\%$ uncertainty, which comes from the observed variation in $(d,p)$ coincidence/singles ratios over different proton angular bins, was propagated into the final \nuc{25}{Mg}/\nuc{26}{Mg} efficiency ratio.
Acceptances for \nuc{25}{Mg} and \nuc{26}{Mg} recoils were determined from Monte Carlo simulations  performed using the NPTOOL  interface to the GEANT4 framework \cite{Matta2016}. The simulations impinged a \nuc{22}{Ne} beam with realistic energy spread and emittance onto the target and generated deuterons and \nuc{26}{Mg} recoils from standard four-momentum conservation in the $({}^{6}\mathrm{Li},d)$ reaction. For the ${}^{25}\mathrm{Mg} + d$ case, the \nuc{26}{Mg} recoils were subsequently broken up into ${}^{25}\mathrm{Mg}^{\mathrm{(g.s.)}}  + n$ assuming an isotropic distribution (only the \nuc{25}{Mg} ground state is energetically accessible for $E_x \leq 11.68$~MeV). The resulting \nuc{25,26}{Mg} recoils, in coincidence with deuteron angles detected in the experiment, were then propagated to the focal plane of the MDM. A first set of acceptance cuts was placed at the location of the $\pm 2^\circ$ slits at the spectrometer entrance. Recoils passing these cuts were then propagated to the end of the spectrometer using a well-characterized RAYTRACE transport code \cite{Pringle1986}. The final set of acceptance cuts was placed at the entrance window to the Oxford detector, which was $\pm 15$~cm and $\pm 3$~cm in the dispersive and non-dispersive planes, respectively. To account for spin dependence, separate simulation runs were performed for $L=0,1,2$ transitions and the weighted average was taken as the final acceptance. 
The acceptances of  \nuc{25}{Mg} and \nuc{26}{Mg} recoils were determined to be $77.8(11)\%$ and $90.8(5)\%$, respectively, resulting in a \nuc{25}{Mg}/\nuc{25}{Mg} efficiency ratio of $0.858(33)$.
} 

\begin{table}
\centering
	\footnotesize
\begin{threeparttable}
	\caption{\label{tab:Tab1}Optical parameters used in FRESCO for the DWBA analysis of $^{22}\mathrm{Ne}({}^6\mathrm{Li}, d){}^{26}\mathrm{Mg}$. All radii except those for the $\alpha + d$ channel are given such that $R_x = r_x  A_T^{1/3}$. For the $\alpha + d$ channel, $R_x=r_x$. }
\begin{tabular}{cccccccc}
\hline
			\multirow{2}{*}{Channel} &
		           $r_c$ & $V_r$ & $r_r$ & $a_r$ & $W_s$ & $r_I$ & $a_I$ \\
		     	  & (fm)  & (MeV) & (fm)  & (fm)  & (MeV) & (fm)  & (fm)  \\
\hline
       $^{22}$Ne+$^6$Li & 1.30  & 117.04& 1.80  & 0.40  & 48.6  & 1.99  & 0.62  \\
       $^{26}$Mg+$d${\tnote{a}} & 1.30  & 79.07 & 1.17  & 0.79  & 2.99  & 1.325 & 0.737 \\
$\alpha$+$d$             & 1.90  & {\tnote{b}}  & 1.90  & 0.65  &       &       &       \\
Final State              & 1.40  & {\tnote{c}}  & 1.40  & 0.70  &       &       &       \\
\hline
\end{tabular}
\begin{tablenotes}
	\item[a]{In addition, the following parameters were used for ${}^{26}\mathrm{Mg} + d$ channel: $W_D=10.51$~MeV, 
$V_{so}=5.88$~MeV, $r_{so}=1.07$~fm, and $a_{so}=0.66$~fm.}
	\item[b]{Adjusted to give the correct $^6$Li binding energy.}
	\item[c]{Adjusted to give the correct final state binding energy.}
\end{tablenotes}
\end{threeparttable}
\end{table}

\begin{figure*}
	\centering
 \hspace*{-1.25in}
 \includegraphics[width=18cm]{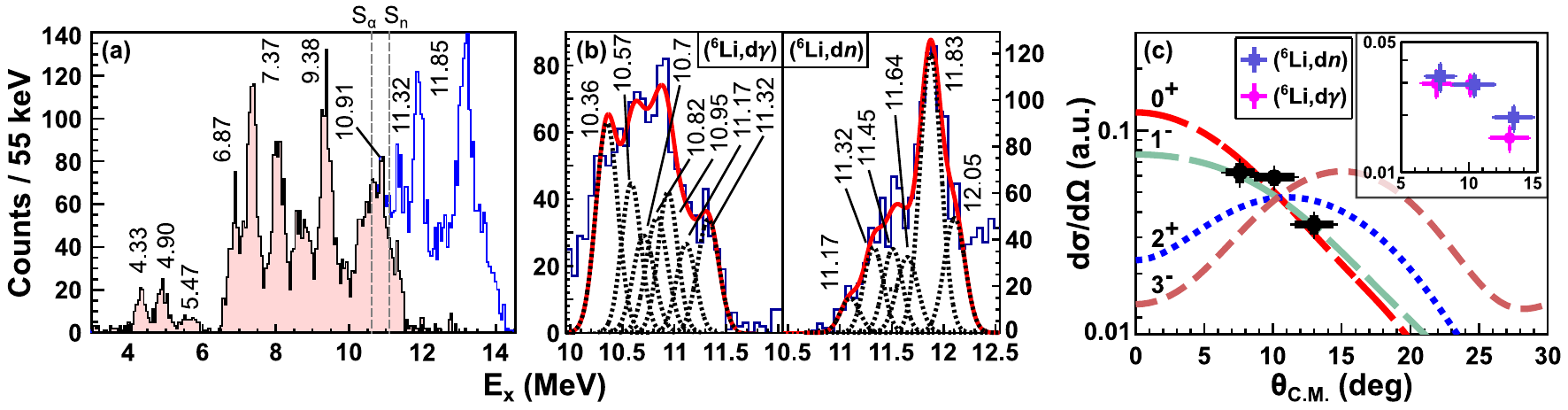}
 \caption{a) $^{26}$Mg excitation energy spectrum measured from the $^6\mathrm{Li}({}^{22}\mathrm{Ne},d){}^{26}$Mg reaction at $\theta_{CM} = 6^{\circ}$--14$^{\circ}$. The red shaded area represents deuterons in coincidence with $^{26}$Mg, while the blue curve is the sum of $^{25}$Mg and $^{26}$Mg coincidences. b) Results of the multi-Gaussian fit explained in the text. Separate fits are shown for $({}^{6}\mathrm{Li}, dn)$ and $({}^{6}\mathrm{Li}, d\gamma)$ as indicated. The black dotted curves represent the individual Gaussian peaks, and the red solid curve is their sum. Nominal energies of each peak included in the fit are enumerated in the figure. c) Angular differential cross section (arbitratily normalized) to the $E_x = 11.32$~MeV state, along with the DWBA calculations explained in the text. The inset shows the individual $^{22}\mathrm{Ne}({}^{6}\mathrm{Li},dn){}^{25}\mathrm{Mg}$ and $^{22}\mathrm{Ne}({}^{6}\mathrm{Li},d\gamma){}^{26}\mathrm{Mg}$ cross sections, as indicated. \label{fig:Fig2}}
 \end{figure*}

Figure~\ref{fig:Fig2}(a) shows the angle-integrated ($\theta_{CM} = 6^{\circ}$--14$^{\circ}$) $^{26}$Mg excitation energy spectrum measured by the annular Si detector. A number of strongly populated states (both bound and unbound) are evident, and the overall features of the spectrum agree well with past measurements \cite{Anantaraman1977, Giesen1993, Talwar2016}. Figure~\ref{fig:Fig2}(c) shows the angular differential cross section (arbitrarily normalized) for the $E_x =11.32$~MeV resonance. Distorted Wave Born Approximation (DWBA) calculations for various $J^\pi$ values are also shown.   The calculations were performed with the FRESCO code \cite{Thompson1988, Thompson2009}, using the optical potential parameters presented in Table~\ref{tab:Tab1}. The optical potential parameters were arrived at by using the SFRESCO minimization routine to best fit the digitized $^{22}\mathrm{Ne}({}^6\mathrm{Li}, d){}^{26}\mathrm{Mg}$ data from Ref.~\cite{Anantaraman1977}, for strongly-populated states with known spin. SFRESCO was used to adjust both the $^6\mathrm{Li} + {}^{22}$Ne optical potentials (starting from those published in Ref.~\cite{Cook1982}) and the $\alpha + {}^{22}$Ne overlap potentials. The  potentials for $d + {}^{26}$Mg and $\alpha + d$ were taken from Refs.~\cite{Daehnick1980, Nishioka1984} and left fixed. For the internal-state ($\alpha$--$d$ system) in the $^{6}$Li nucleus, a relative $2S$ state was assumed \cite{Anantaraman1977}. For the final state in the $^{26}$Mg nucleus, the number of radial nodes was fixed by the harmonic oscillator energy conservation relation, assuming an $(sd)^4$ configuration for the positive parity states and an $(sd)^3(fp)$ configuration for the negative parity states. 
For resonance states,  $J^\pi$ radial wave functions, computed using the final-state $\alpha + {}^{22}\mathrm{Ne}$ potential, were used.

\subsection{11.32 MeV Resonance}\label{sec:1132}
A peak at $E_x$ = 11.32 MeV is clearly present in the summed ${}^{25}\mathrm{Mg} + {}^{26}\mathrm{Mg}$ excitation energy spectrum, with approximately equal numbers of \nuc{25,26}{Mg} coincidences. As shown in Figure~\ref{fig:Fig2}(c), the shape of our summed ${}^{25}\mathrm{Mg} + {}^{26}\mathrm{Mg}$ angular distribution is consistent with the $J^\pi = (0^{+},~1^-)$ calculations and not consistent with  $J^\pi = (2^{+},~3^-)$. The angular distributions for the neutron and $\gamma$-ray decay channels are statistically identical in shape, as demonstrated in the inset of Figure~\ref{fig:Fig2}(c). This strongly suggests that the resonances observed in independent $(\alpha,n)$ and $(\alpha,\gamma)$ direct measurements are indeed the same state and not a doublet as suggested in Refs.~\cite{Longland2012, Adsley2017}. Owing to the lack of reliable beam-on-target normalization, we were unable to extract absolute $\alpha$ spectroscopic factors for the $E_x=11.32$~MeV state. Instead, we determine relative spectroscopic factors by normalizing to the published direct-measurement $(\alpha,\gamma)$ strength, combined with the presently-reported $\Gamma_{n}/\Gamma_{\gamma}$. For the direct-measurement strengths, we take the weighted average of Refs.~\cite{Wolke1989,Hunt2019}, $37(4)~\mu$eV. From the spectroscopic factors obtained using this procedure, we calculate spin-dependent values of $\Gamma_{\alpha}$ using the prescription of Ref.~\cite{Thompson2009}. The results are presented in Table~\ref{tab:Table2}, assuming both $0^+$ and $1^-$ assignments.

\revised{
To extract the $\Gamma_n / \Gamma_\gamma$ for the $E_x = 11.32$~MeV resonance, we performed a multiple-Gaussian fit to the region around  $E_x = 11.32$~MeV in the individual ${}^{25}$Mg- and ${}^{26}$Mg-gated spectra. The amplitude of each Gaussian was allowed to vary freely, and the central energies and widths were restricted based on existing experimental information and knowledge of the experimental setup, as described below.
In the fit, we included all states previously observed in $({}^{6}\mathrm{Li},d)$ by either of Refs.~\cite{Giesen1993, Talwar2016}. The nominal central energies of each of these peaks are enumerated in Figure~\ref{fig:Fig2}(b). To account for uncertainties in the peak energy,  the centroid of each Gaussian was allowed to vary by $\pm 10$~keV from the nominal---with the exception of the peak at ${\sim} 11.17$~MeV. Because multiple natural-parity states have been identified in the region around 11.17~MeV \cite{Massimi2017}, and furthermore because of the controversy between Ref.~\cite{Giesen1993} and Ref.~\cite{Talwar2016} concerning the presence of a low-spin, strong $\alpha$ cluster state in this region, we allowed for a greater uncertainty in the central energy of the peak in this region. Specifically, we allowed the central energy of this peak to vary freely between 11.1 and 11.25~MeV. Because of the astrophysical interest of this state, we also extracted an upper limit on its spectroscopic factor, as explained in Section~\ref{sec:11171}.
The width of each peak was allowed to vary by $\pm 11.8$~keV FWHM ($\sigma= \pm 5$ keV) from the nominal resolution of $230$~keV FWHM. The resolution was determined from a GEANT4 Monte Carlo simulation that included the the effects of target thickness, beam emittance and energy spread, and Si energy and angular resolution. The accuracy of the simulation was verified by comparing the results to experimental data for strongly populated states.
} 

The results of the fitting procedure are shown in Figure~\ref{fig:Fig2}(b), overlayed with the present experimental data. The total number of counts in the \nuc{25}Mg- and \nuc{26}{Mg}-gated, $1.32$~MeV peaks were extracted from the areas under the respective \nuc{25}Mg- and \nuc{26}{Mg}-gated Gaussians centered at 11.32 MeV. Their efficiency-corrected ratio resulted in a branching ratio $\Gamma_n/\Gamma_\gamma = 1.14(26)$.
Normalizing to $\omega\gamma_{(\alpha,\gamma)} = 37(4)~\mu$eV,  we obtain a new $^{22}\mathrm{Ne}(\alpha,n){}^{25}\mathrm{Mg}$ strength of $\omega\gamma_{(\alpha,n)} = 42(11)~\mu$eV. While this strength agrees within $2\sigma$ with the direct measurement of Harms \emph{et al}., $83(24)~\mu$eV \cite{Harms1991}, it disagrees by more than $2\sigma$ with all other direct measurements. In particular it disagrees by $5.0\sigma$ with the most recently published result of Jaeger \emph{et al.}, $118(11)~\mu$eV \cite{Jaeger2001}, and by $3.1\sigma$ with the inflated weighted average of $140(30)~\mu$eV calculated by  Longland \emph{et al}.\ \cite{Longland2012}.

The  implications of the presently-established $(\alpha,n)$ resonance strength are fully realized when it is combined with the $\Gamma_\alpha$ concurrently reported in a sub-Coulomb ${}^{6}\mathrm{Li}({}^{22}\mathrm{Ne},\linebreak[1] d){}^{26}\mathrm{Mg}$ and ${}^{7}\mathrm{Li}({}^{22}\mathrm{Ne},\linebreak[1] t){}^{26}\mathrm{Mg}$ study by Jayatissa \emph{et al}.\ \cite{Jayatissa2019}. As explained in that work, the present $\Gamma_n/\Gamma_\gamma$ can be combined with the direct-measurement $(\alpha,\gamma)$ strength of $37(4)~\mu$eV to calculate $\alpha$ partial widths of $79(13)$, $26(4)$, and $16(3)~\mu$eV for respective $J=0,$ $1$, and $2$ spin assignments. The respective levels of agreement with the sub-Coulomb results are $1.1\sigma$, $2.8\sigma$, and $5.0\sigma$. This strongly suggests a $J^\pi=0^+$ assignment for this state, although the $1^-$ assignment cannot be conclusively ruled out. These assignments agree with the present angular distributions, shown in Figure~\ref{fig:Fig2}, which are consistent with $J=(0,1)$.

A similar argument to the one above suggests that the $\omega \gamma_{(\alpha,n)}^{(11.32)}$ reported in direct measurements are seriously overestimated. 
\revised{
If we instead take $(2J+1)\Gamma_\alpha = \omega\gamma_{(\alpha,n)}+\omega\gamma_{(\alpha,\gamma)}$, with $\omega\gamma_{(\alpha,n)}$ and $\omega\gamma_{(\alpha,\gamma)}$ both from direct measurements, even the $J^\pi = 0^+$ results disagree substantially with  Ref.~\cite{Jayatissa2019}.
} 
For example, using $\omega\gamma_{(\alpha,n)}= 118(11)~\mu$eV \cite{Jaeger2001} results in $\Gamma_\alpha^{(J=0)} = 155(12)~\mu$eV, a $5.9\sigma$ discrepancy. Similarly, using the inflated weighted average presented by Longland \emph{et al}., $\omega\gamma_{(\alpha,n)}= 140(30)~\mu$eV, results in $\Gamma_\alpha^{(J=0)} = 177(30)~\mu$eV, a $3.6\sigma$ discrepancy.

\begin{table*}
	\centering
	\footnotesize
	\begin{threeparttable}
		\caption{\label{tab:Table2}Resonance parameters determined for the $^{26}$Mg $11.32$~MeV state and natural parity states between the neutron threshold and 11.25~MeV identified in Refs.~\cite{Massimi2012, Massimi2017}. All upper limits are quoted at $90\%$ CL, and uncertainties in parentheses are quoted at $1\sigma$ ($68\%$ CL).}
\begin{tabular}{ccccccccc}
\hline
	$E_x$ & $E_{cm}$ & \multirow{2}[0]{*}{$J^\pi$} & \multirow{2}[0]{*}{${\Gamma_n}/{\Gamma_\gamma}$} & \multirow{2}[0]{*}{$S_\alpha$} & $\Gamma_\alpha$ & $\omega\gamma_{tot}$ & $\omega\gamma_{(\alpha,\gamma)}$ & $\omega\gamma_{(\alpha,n)}$ \\
	(MeV) & (MeV) &       &       &       & (eV)  & (eV)  & (eV)  & (eV) \\
\hline
	11.112 & 0.497 & $2^+$\tnote{a} & 1530\tnote{a} & $<0.025$\tnote{b} & $<2.2\ee{-10}$\tnote{b} & $<1.1\ee{-9}$ & $<7.1\ee{-13}$ & $<1.1\ee{-9}$ \\
	&      &       &                                & $<0.043$\tnote{c} & $<4.3\ee{-10}$\tnote{c} & $<2.1\ee{-9}$ & $<1.4\ee{-12}$ & $<2.1\ee{-9}$ \\
	11.163 & 0.548 & $2^+$\tnote{a} & 1900\tnote{a} & $<0.025$\tnote{b} & $<2.7\ee{-9}$\tnote{b}  & $<1.4\ee{-8}$ & $<7.2\ee{-12}$ & $<1.4\ee{-8}$ \\
	&      &       &                                & $<0.043$\tnote{c} & $<5.2\ee{-9}$\tnote{c}  & $<2.6\ee{-8}$ & $<1.4\ee{-11}$ & $<2.6\ee{-8}$ \\
	11.169 & 0.554 & $3^-$\tnote{a} & 588\tnote{a}  & $<0.024$\tnote{b} & $<4.4\ee{-10}$\tnote{b} & $<3.1\ee{-9}$ & $<5.2\ee{-12}$ & $<3.1\ee{-9}$ \\
	&      &       &                                & $<0.041$\tnote{c} & $<1.1\ee{-11}$\tnote{c} & $<5.9\ee{-9}$ & $<1.0\ee{-11}$ & $<5.9\ee{-9}$ \\
	11.171 & 0.556 & $2^+$\tnote{a} & 0.2\tnote{a}  & $<0.15$\tnote{b}  & $<1.9\ee{-8}$\tnote{b}  & $<9.6\ee{-8}$ & $<8.0\ee{-8}$  & $<1.6\ee{-8}$ \\
	&      &       &                                & $<0.26$\tnote{c}  & $<3.7\ee{-8}$\tnote{c}  & $<1.9\ee{-7}$ & $<1.6\ee{-7}$  & $<3.1\ee{-8}$ \\
	11.318 & 0.703 & $0^+$ & $1.14(26)$ & $0.31(5)$\tnote{d} & $7.9(13)\ee{-5}$\tnote{d} & $7.9(13)\ee{-5}$ & $3.7(4)\ee{-5}$\tnote{e} & $4.2(11)\ee{-5}$\tnote{d} \\
	             &             & $1^-$  & $1.14(26)$ & $0.18(3)$\tnote{d} & $2.6(4)\ee{-5}$\tnote{d}  & $7.9(13)\ee{-5}$ & $3.7(4)\ee{-5}$\tnote{e} & $4.2(11)\ee{-5}$\tnote{d} \\
\hline
\end{tabular}%
		\begin{tablenotes}
			\item[a]{Adopted from Refs.~\cite{Massimi2012, Massimi2017}.}
			\item[b]{Normalized to $S_\alpha^{(E_x=11.32)}$, assuming $J^\pi=0^+$. }
			\item[c]{Normalized to $S_\alpha^{(E_x=11.32)}$, assuming $J^\pi=1^-$. }
			\item[d]{Normalized to $\omega\gamma_{(\alpha, \gamma)} = 37(4)~\mu$eV and $\Gamma_{n}/\Gamma_{\gamma}=1.14(26)$.}
			\item[e]{Weighted average of Refs.~\cite{Hunt2019, Wolke1989}.}
		\end{tablenotes}
	\end{threeparttable}
\end{table*}

\subsection{Resonances Below 11.32 MeV}
\label{sec:11171}

\revised{
Between the neutron threshold and $E_x = 11.32$~MeV, we observe no clearly resolved peak in the $^{25}$Mg-gated, $^{26}$Mg-gated, or summed spectrum. However, the data are not well described by a fit that does not include a peak in this region. For a conservative treatment of states in this region, we use the results of the multiple-Gaussian fit described in Section~\ref{sec:1132} to extract an upper limit on the cross section, and hence $S_\alpha$, for a hypothetical state in this energy regime.
Specifically, we use the amplitude of the ${\sim} 11.17$~MeV peak in the $^{25}$Mg-gated spectrum to set a $90\%$ confidence level (CL) upper limit on the ${}^{22}\mathrm{Ne}({}^6\mathrm{Li}, dn){}^{25}\mathrm{Mg}$ cross section for \emph{any one} state between 11.17--11.25~MeV. We then use this cross section to calculate  upper limits on the $({}^6\mathrm{Li}, dn)$ cross section for each of the four natural-parity states identified in Ref~\cite{Massimi2017} ($E_x = 11.112$, $11.163$, $11.169$, and $11.171~$MeV). For this, we assume that $100\%$ of the observed upper-limit strength goes into each state individually. This results in a conservative upper limit because if the strength were shared between one or more states, the resulting cross sections would be lower for each.
} 


From the upper limits on the ${}^{22}\mathrm{Ne}({}^6\mathrm{Li}, dn){}^{25}\mathrm{Mg}$ cross sections, we calculate upper limits on $S_\alpha \Gamma_{n} / \Gamma$ by normalizing to the presently-observed $S_\alpha$ for the $E_x=11.32$~MeV resonance, which is in turn normalized to $\omega\gamma^{\mathrm(11.32)}_{(\alpha,\gamma)}$. We take the $J^\pi$ for each of the four lower-energy resonances from Refs.~\cite{Massimi2012, Massimi2017} and do separate normalizations for both the $0^+$ and $1^-$ assignments to the $E_x=11.32$~MeV resonance. We convert these into limits on $S_\alpha$ by multiplying by $\Gamma/\Gamma_n$, taking $\Gamma_n$ and $\Gamma_\gamma$ from Ref.~\cite{Massimi2017} (and assuming $\Gamma \simeq \Gamma_n + \Gamma_\gamma$).
%
For the $E_x=11.171$~MeV resonance, which has a significant $\gamma$-ray decay branch and a large uncertainty on the neutron width, we adopt $\Gamma_n/\Gamma = 1/6$ 
for these calculations, i.e.\ the smallest value consistent with Ref.~\cite{Massimi2017}. This gives a conservative upper limit on $S_\alpha$ since a smaller $\Gamma_n/\Gamma_\gamma$ results in a larger $S_\alpha$ using the present procedure. 
The resulting upper limits are presented in Table~\ref{tab:Table2}, along with a summary of the resonance parameters adopted for the $E_x = 11.32$~MeV state.

The $\gamma$-decaying resonance at $E_x=11.171$~MeV  is the likely candidate for the strong  $\alpha$ cluster state with $J^\pi = (1^-, 2^+)$ reported by Talwar \emph{et al}.\ \cite{Talwar2016}. Our upper limit on the spectroscopic factor for this state is substantially below that reported in Ref.~\cite{Talwar2016}, even when the latter is re-normalized to the present $\omega\gamma_{tot}^{(11.32)}$. 
An even more stringent upper limit of $\Gamma_\alpha^{(J=2)} < 1.3\ee{-11}~$eV is reported in the concurrent sub-Coulomb study, Ref.~\cite{Jayatissa2019}.
A possible reason for the discrepancy between Ref.~\cite{Talwar2016} and the present experiment (along with Refs.~\cite{Giesen1993, Jayatissa2019}) is that the state observed in Ref.~\cite{Talwar2016} is actually a higher-spin state ($J \geq 3$) and thus more likely to be populated with their beam energy of $E^{(\mathrm{lab})}_{{}^6\mathrm{Li}} = 82.3$~MeV. This would also be consistent with a very recent GAMMASPHERE measurement which observed $\gamma$-ray decay from a state at $E_x=11.171$~MeV to a $4^+$ level, indicating a spin ranging from $2$--$6$ \cite{Lotay2019}.

%


\section{Astrophysical Implications}

\begin{figure}
\centering
	\includegraphics[width=8cm]{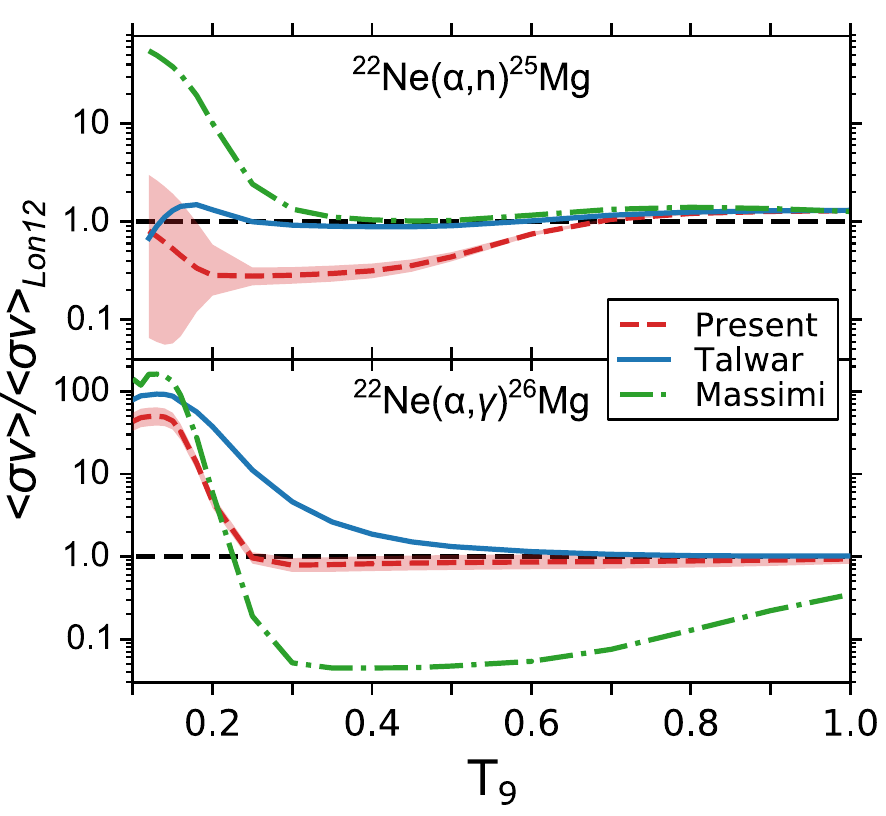}
	\caption{Updated ${}^{22}\mathrm{Ne}(\alpha,n){}^{25}\mathrm{Mg}$ (top) and ${}^{22}\mathrm{Ne}(\alpha,\gamma){}^{26}\mathrm{Mg}$ (bottom) reaction rates, presented as a ratio to the recommended rates given by Longland \emph{et al}., Ref.~\cite{Longland2012}. The red dashed line represents our recommended rate, while the surrounding band represents the extent between our low and high rates. Also included for comparison are the recommended rates (also given as ratios to Ref.~\cite{Longland2012}) from Talwar \emph{et al}.~\cite{Talwar2016} (solid blue curve labeled ``Talwar'' in the legend) and  the upper limits presented in Massimi \emph{et al}.~\cite{Massimi2017} (green dot-dashed curve labeled ``Massimi'').
\label{fig:Rates}
}
\end{figure}

\begin{figure}
\centering
	\includegraphics[width=8cm]{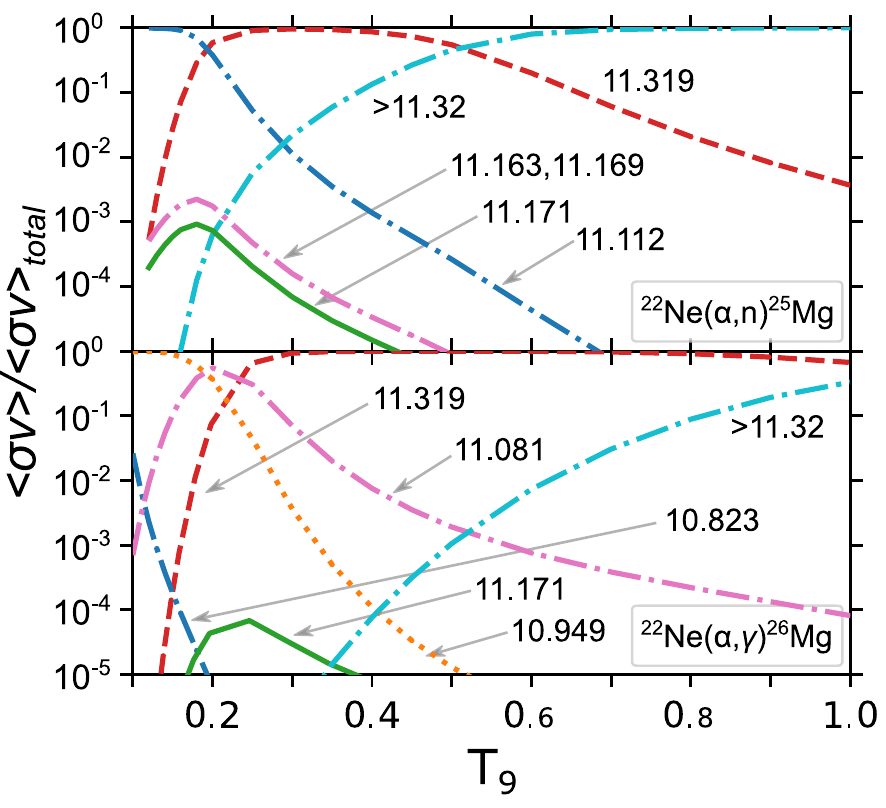}
	\caption{Fractional resonance contributions of individual resonances to the median ${}^{22}\mathrm{Ne}(\alpha,n){}^{25}\mathrm{Mg}$ (top) and ${}^{22}\mathrm{Ne}(\alpha,\gamma){}^{26}\mathrm{Mg}$ (bottom)  rates.		
	\label{fig:RatesCompare}
}
\end{figure}

\begin{table}
\centering
\footnotesize
\begin{threeparttable}
\caption{Resonance properties adpoted for the Monte Carlo rate calculations.\label{tab:RateProperties} }
\begin{tabular}{cccc}
\hline
$E_x$ & \multirow{2}{*}{$J^\pi$} & \multirow{2}{*}{$\Gamma_n/\Gamma_\gamma$} & $\Gamma_\alpha$ \\
(MeV) & & & (eV) \\ \hline
10.8226(30)\tnote{a} & $2^+$ \tnote{a} & $0$                   & $2.1 \pm 0.3\mathrm{(stat)} \pm 0.4 (\mathrm{sys})\ee {-22}$\tnote{d,g}\\
10.9491(8)\tnote{a}  & $1^-$ \tnote{a} & $0$                   & $3.0 \pm 0.3\mathrm{(stat)}{}^{+0.8}_{-0.6}(\mathrm{sys})\ee{-14}$\tnote{d,g} \\
11.0809(40)\tnote{a} & $2^+$ \tnote{a} & $0$                   & $5.7 \pm 0.7\mathrm{(stat)}{}^{+1.4}_{-1.2}(\mathrm{sys})\ee{-11}$\tnote{d,g} \\
11.112(6)\tnote{b}   & $2^+$\tnote{b} & $1530(67)$\tnote{b,e}  & $< 2.2\ee{-10}$\tnote{f} \\
11.163(2)\tnote{b}   & $2^+$\tnote{b} & $1896(137)$\tnote{b,e} & $< 1.3\ee{-11}$\tnote{d,h} \\
11.169(1)\tnote{b}   & $3^-$\tnote{b} & $588(36)$\tnote{b,e}   & $< 1.3\ee{-11}$\tnote{d,h} \\
11.171(1)\tnote{b}   & $2^+$\tnote{b} & $0.2$--$6$\tnote{b}    & $< 1.3\ee{-11}$\tnote{d,h} \\
11.3195(25)\tnote{c} & $0^+$\tnote{d} & $1.14(26)$\tnote{f}    & $7.9(13) \ee{-5}$\tnote{f} \\
\multirow{2}{*}{${>} 11.32$} & \multicolumn{3}{p{5.8cm}}{Resonance strengths and energies adopted from Ref.~\cite{Jaeger2001} for $(\alpha,n)$ and Ref.~\cite{Wolke1989} for $(\alpha,\gamma)$.} \\
\hline
\end{tabular}
		\begin{tablenotes}
			\item[a]{Adopted from Ref.~\cite{Lotay2019}.}
			\item[b]{Adopted from Ref.~\cite{Massimi2017}.}
			\item[c]{Adopted from Ref.~\cite{Hunt2019}.}
			\item[d]{Adopted from Ref.~\cite{Jayatissa2019}.}
			\item[e]{Treated as negligable in the $(\alpha,\gamma)$ rate calculation.}
			\item[f]{Adopted from the present work.}
			\item[g]{A common (correlated) systematic uncertainty of ${}^{+25\%}_{-21\%}$ has been generated for each of these resonances.}
			\item[h]{See text for a detailed explanation.}
		\end{tablenotes}
	\end{threeparttable}
\end{table}

To investigate the impact of our measurements on the stellar ${}^{22}\mathrm{Ne}(\alpha,n){}^{25}\mathrm{Mg}$ and ${}^{22}\mathrm{Ne}(\alpha,\gamma){}^{26}\mathrm{Mg}$ reaction rates, we have calculated low, recommended, and high rates using a Monte-Carlo procedure similar to the one described in Ref.~\cite{Longland2010}. This procedure accounts for uncertainties in nuclear physics quantities by treating them as probability density functions (PDFs), randomly sampling each quantity from its presumed PDF over many trials (here we use $N=50,000$). The resulting distribution of reaction rates is then analyzed to extract the low ($15^{th}$ percentile), recommended ($50^{th}$ percentile), and high ($85^{th}$ percentile) rates.

The $(\alpha, n)$ calculations include contributions from the five resonances treated in the present work, along with resonances above $E_x = 11.32$~MeV. The $(\alpha, \gamma)$ calculations include the four resonances observed in Ref.~\cite{Jayatissa2019}, along with the $E_x=11.171$~MeV resonance and resonances above $11.32$~MeV. A detailed list of the adopted excitation energies, $n/\gamma$ branching ratios, and $\alpha$ partial widths, including the associated uncertainties, is given in Table~\ref{tab:RateProperties}. The states at $E_x = 11.163$, $11.169$, and $11.171$~MeV require special consideration, as the $11.163$ and $11.169$ states were not treated in Ref.~\cite{Jayatissa2019}. However, because they are so close in energy to the $11.171$~MeV state, the experimental cross section limit of $0.8~\mu$b/sr is actually the limit for the \emph{total} population of all three states. For a conservative treatment, we adopt individual upper limits of $\Gamma_\alpha < 1.3\ee{-11}$~eV for each of these three states. This limit is only correct for $J=2$ states; however the upper limit for the $3^-$ state at $E_x=11.169$~MeV  would be even lower, given the same experimental cross section. Hence the application of the $J = 2$ limit to this state is again conservative. 

\begin{table}
\centering
\footnotesize
\caption{Monte Carlo rates calculated for the ${}^{22}\mathrm{Ne}(\alpha,n){}^{25}\mathrm{Mg}$ and ${}^{22}\mathrm{Ne}(\alpha,\gamma){}^{26}\mathrm{Mg}$ reactions. The rates are presented as $\log_{10} \left( N_A \left\langle \sigma v \right\rangle \right)$, in units of cm$^3$/mol/s. \label{tab:Rates} }
\begin{tabular}{cccc|ccc}
 & \multicolumn{3}{c}{${}^{22}\mathrm{Ne}(\alpha,n){}^{25}\mathrm{Mg}$} &  \multicolumn{3}{c}{${}^{22}\mathrm{Ne}(\alpha,\gamma){}^{26}\mathrm{Mg}$}\\ 
\hline
 $T_9$    & Low & Median & High & Low & Median & High \\
\hline
0.10 & -- & -- & --             & -24.12 & -24.01 & -23.90   \\
0.11 & -- & -- & --             & -22.66 & -22.54 & -22.44   \\
0.12 & -25.96 & -24.89 & -24.32 & -21.44 & -21.32 & -21.21   \\
0.13 & -24.39 & -23.33 & -22.76 & -20.40 & -20.29 & -20.18   \\
0.14 & -23.02 & -22.00 & -21.43 & -19.51 & -19.39 & -19.29   \\
0.15 & -21.79 & -20.84 & -20.28 & -18.72 & -18.61 & -18.51   \\
0.16 & -20.63 & -19.82 & -19.27 & -18.01 & -17.91 & -17.82   \\
0.18 & -18.48 & -18.04 & -17.57 & -16.77 & -16.68 & -16.60   \\
0.20 & -16.66 & -16.45 & -16.14 & -15.68 & -15.60 & -15.53   \\
0.25 & -13.34 & -13.24 & -13.15 & -13.21 & -13.14 & -13.07   \\
0.30 & -11.11 & -11.02 & -10.94 & -11.13 & -11.05 & -10.97   \\
0.35 & -9.50 & -9.42 & -9.34    & -9.57 & -9.49 & -9.40      \\
0.40 & -8.28 & -8.20 & -8.13    & -8.40 & -8.31 & -8.22      \\
0.45 & -7.28 & -7.21 & -7.15    & -7.49 & -7.41 & -7.31      \\
0.50 & -6.40 & -6.36 & -6.31    & -6.77 & -6.69 & -6.59      \\
0.60 & -4.86 & -4.84 & -4.82    & -5.71 & -5.62 & -5.53      \\
0.70 & -3.59 & -3.58 & -3.56    & -4.95 & -4.86 & -4.78      \\
0.80 & -2.57 & -2.56 & -2.55    & -4.37 & -4.29 & -4.21      \\
0.90 & -1.75 & -1.74 & -1.72    & -3.89 & -3.82 & -3.75      \\
1 .00& -1.08 & -1.06 & -1.05    & -3.47 & -3.42 & -3.35      \\
\hline
\end{tabular}
\end{table}

\revised{
In Figure~\ref{fig:Rates}, the resulting low, recommended, and high rates are plotted as a ratio to the recommended rates given by Longland \emph{et al}.\ in Ref.~\cite{Longland2012}.
For comparison, we also show the ratio-to-Longland of the recommended rates given by \etal{Talwar} \cite{Talwar2016} and the upper limits of \etal{Massimi} \cite{Massimi2017}. Numerical values of our low, recommended, and high rates are also given in Table~\ref{tab:Rates}.
Figure~\ref{fig:RatesCompare} shows the individual contributions of each resonance to the overall rate. We find that for the ${}^{22}\mathrm{Ne}(\alpha, n){}^{25}\mathrm{Mg}$ reaction, the $E_x=11.32$~MeV resonance completely dominates the total rate in the  temperature regime from ${\sim} 0.2$--$0.4$~GK,  where ${}^{22}\mathrm{Ne}(\alpha,\linebreak[1] n){}^{25}\mathrm{Mg}$ is thought to be the primary neutron source in the He-core burning phases of AGB stars. The presently-established $(\alpha,n)$ strength of this resonance results in a recommended rate that is up to a factor  ${\sim} 3$ lower than Refs.~\cite{Talwar2016,Longland2012,Massimi2017} in the crucial temperature regime from 0.2--0.4~GK.
For $T>0.4$~GK, resonances above $E_x = 11.32$~MeV begin to contribute to the overall $(\alpha,n)$ rate, eventually becoming dominant for $T>0.6$~GK. Below $0.2$~GK, the $E_x = 11.112$~MeV resonance is potentially dominant. This resonance still carries large uncertainties on its $\alpha$ partial width, leading to the large uncertainty band for the overall rate in this temperature region. As a result, a crucial focus of future measurements should be establishing tighter limits on the $\alpha$ width of this resonance.
We note additionally that the resonances at $E_x = 11.163,$ $11.169$, and $11.171$ are nearly inconsequential to the overall rate---with a fractional contribution of ${\sim} 10^{-3}$ or less across the entire relevant temperature regime for the $s$-process. This conclusion is the result of the stringent upper limits set on the $\alpha$ partial width in Ref.~\cite{Jayatissa2019}.


 For the ${}^{22}\mathrm{Ne}(\alpha,\linebreak[1] \gamma){}^{26}\mathrm{Mg}$ reaction, the total rate is dominated below ${\sim} 0.2$~GK by the $10.823$ and $10.949$ resonances---whose strengths are now well characterized by Ref.~\cite{Jayatissa2019}. Between ${\sim} 0.2$--$0.25$~GK the $11.081$~keV resonance dominates, and again the strength of this resonance is now well characterized by Ref.~\cite{Jayatissa2019}.
Above ${\sim} 0.25$~GK, the $E_x = 11.32$~MeV resonance, with a well-established strength of 37(4)~$\mu$eV, dominates. As with the $(\alpha, n)$ rate, the $E_x = 11.171$~MeV resonance is completely inconsequential---less than a  $10^{-4}$ fractional contribution to the total rate over all temperatures. As a result, our recommended rate is substantially below Talwar \emph{et al.}\ (maximum factor ${\sim} 10$) up to ${\sim} 0.6$~GK. In contrast, the present rate is significantly above the \etal{Longland} recommended rate below 0.25~GK (maximum factor ${\sim} 45$). Below 0.2~GK, this is primarily the result of \etal{Longland} using an upper limit of $\Gamma_\alpha < 2.9\ee{-15}$~eV for the 10.949~MeV resonance, adopted from Ref.~\cite{Ulgade2007}. Subsequent to that publication, strengths at least an order of magnitude larger than this limit have been established by both Ref.~\cite{Talwar2016} and Ref.~\cite{Jayatissa2019}, leading to the much larger $(\alpha,\gamma)$ rate in this temperature regime presented both in the present work and in Ref.~\cite{Talwar2016}. Between 0.2--0.25~GK, the higher rate is mainly the result of Ref.~\cite{Longland2012} neglecting the 11.081~MeV resonance, whose contribution again was not established until Refs.~\cite{Talwar2016, Jayatissa2019}.
%

We emphasize that the rates presented here, and the corresponding conclusions, are dependent on the spin assignments adopted in Table~\ref{tab:RateProperties}. More conservative rate calculations, which account for uncertainties in the spin, can be found in Ref.~\cite{Jayatissa2019}.  The overall conclusions are not subtantially different between the two treatments.
} 

\section{Summary}

In summary, we have measured the $^{22}\mathrm{Ne}({}^6\mathrm{Li}, d){}^{26}\mathrm{Mg}$ reaction in inverse kinematics, using a $^{22}\mathrm{Ne}$ beam with an energy of $154$~MeV. We detected both the outgoing $d$ and $^{25,26}$Mg in coincidence, which gives sensitivity to $n/\gamma$ branching ratios through recoil tagging. For the key $s$-process resonance at $E_x=11.32$~MeV, we find that $\Gamma_n/\Gamma_{\gamma} = 1.14(26)$, roughly a factor $3$ below the ratio established from direct measurements. Normalizing to the well-known $^{22}\mathrm{Ne}(\alpha, \gamma){}^{26}\mathrm{Mg}$ resonance strength, we establish a new $^{22}\mathrm{Ne}(\alpha, n){}^{25}\mathrm{Mg}$ strength of $42(11)~\mu$eV for the $E_x = 11.32$~MeV resonance.  The angular distributions for this state are consistent with $J^\pi=(0^+,1^-)$ spin-parity assignments, which agrees with Ref.~\cite{Jayatissa2019}.
We note that the presently-established strength is independent of spin assignments or optical model calculations and depends only on $\Gamma_n/\Gamma_\gamma$ and the well-established $\omega\gamma_{(\alpha,\gamma)}^{(11.32)}$.

For neutron-unbound resonances below $E_x = 11.32$~MeV, we have determined upper limits on $\omega\gamma_{(\alpha,n)}$ and $\omega\gamma_{(\alpha,\gamma)}$ by combining an analysis of our ${}^{22}\mathrm{Ne}({}^6\mathrm{Li}, dn){}^{25}\mathrm{Mg}$ specrum with the $\Gamma_n$ and $\Gamma_\gamma$ reported in Ref.~\cite{Massimi2017}. In all cases, our upper limits on the $(\alpha,n)$ strength are below the direct-measurement limit of $6\ee{-8}$~eV reported in Ref.~\cite{Jaeger2001}. However, more stringent limits for the resonances between $E_x = 11.163$--$11.171$~MeV are concurrently set in Ref.~\cite{Jayatissa2019}. As a result we adopt these more restrictive limits for subsequent rate calculations.

\revised{
The presently-established, reduced strength of the $E_x = \linebreak[1] 11.32$~MeV resonance results in a recommended $(\alpha,n)$ rate, calculated using a modern Monte Carlo procedure, that is significantly below Refs.~\cite{Talwar2016, Longland2012, Massimi2017} in the crucial temperature regime from $0.2$--$0.4$~GK. }
The resulting decrease in the neutron flux is expected to reduce predicted $s$-process abundances for elements above mass ${\sim} 60$; however, this is likely to be mitigated by the similar decrease in the $\mathrm{Ne}(\alpha, \gamma){}^{26}\mathrm{Mg}$ rate in the same temperature regime.
Detailed calculations evaluating the impact of the present changes in the recommended $\mathrm{Ne}(\alpha, n){}^{25}\mathrm{Mg}$ and $\mathrm{Ne}(\alpha, \gamma){}^{26}\mathrm{Mg}$ rates on $s$-process abundances will be presented in a forthcoming publication.

Taking the present results together with Ref.~\cite{Jayatissa2019}, we highlight two major, outstanding uncertainties surrounding the $^{22}\mathrm{Ne}(\alpha, \linebreak[1] n){}^{25}\mathrm{Mg}$ reaction rate. First, the $\alpha$ partial width of the $2^+$, $E_x =11.112$~MeV resonance reported by Massimi \emph{et al.}\ \cite{Massimi2017} is not well constrained. This resonance potentially dominates the overall rate below ${\sim} 0.2$~GK due to its low energy---only $19$~keV above the neutron threshold. As a result, the total reaction is not well constrained at low temperatures. Second, the substanial discrepancy between the present $\omega\gamma_{(\alpha,n)}^{(11.32)}$ and direct measurements clearly highlights a need for future direct $(\alpha, n)$ measurements, or complimentary indirect studies of this key resonance. In particular, direct measurements that reduce room background, either by being performed underground or above ground using inverse kinematics and a recoil separator, are particularly welcomed.

\section*{Acknowledgements}
We express our thanks to the technical staff at the Texas A\&M University Cyclotron Institute. Financial support for this work was provided by the US Department of Energy, Office of Science, award No.\ DE-FG02-93ER4077; the US National Nuclear Security Administration, award No.\ DE-NA0003841, and the Nuclear Solutions Institute at Texas A\&M University. J.A.T., W.N.C., and G.L.\ acknowledge support from the Science and Technology Facilities Council (U.K.) Grant No.\ ST/L005743/1.  The ${}^{6}\mathrm{LiF}$ targets were provided by the Center for Accelerator Target Science at Argonne National Laboratory.


%

\end{document}